# Reconfigurable fractional microwave signal processor based on a microcomb

Mengxi Tan,[1] Xingyuan Xu,[1] Jiayang Wu,[1] Thach G. Nguyen,[2] Sai T. Chu,[3] Brent E. Little,[4] Roberto Morandotti,[5,6,7] Arnan Mitchell,[2] and David J. Moss[1,*]

[1]*Centre for Micro-Photonics, Swinburne University of Technology, Hawthorn, VIC 3122, Australia*
[2]*School of Engineering, RMIT University, Melbourne, VIC 3001, Australia*
[3]*Department of Physics and Material Science, City University of Hong Kong, Tat Chee Avenue, Hong Kong, China.*
[4]*Xi'an Institute of Optics and Precision Mechanics Precision Mechanics of CAS, Xi'an, China.*
[5]*INRS-Énergie, Matériaux et Télécommunications, 1650 Boulevard Lionel-Boulet, Varennes, Québec, J3X 1S2, Canada.*
[6]*ITMO University, St. Petersburg, Russia.*
[7]*Institute of Fundamental and Frontier Sciences, University of Electronic Science and Technology of China, Chengdu 610054, China.*
*dmoss@swin.edu.au

*Abstract*—We propose and demonstrate reconfigurable fractional microwave signal processing based on an integrated Kerr optical microcomb. We achieve two forms of microwave signal processing functions – a fractional Hilbert transform as well as a fractional differentiator. For the Hilbert transform we demonstrate a phase shift of 45 degrees – half that of a full Hilbert transform, while for the differentiator we achieve square-root differentiation. For both, we achieve high resolution over a broad bandwidth of 17 GHz with a phase deviation of less than 5° within the achieved passband. This performance in both the frequency and time domains demonstrates the versatility and power of micro-combs as a basis for high performance microwave signal processing.

*Keywords—Microwave photonics, signal processor, micro-ring resonators.*

## I. Introduction

Signal processors, such as Hilbert transformers (HT) and differentiators, are one of the basic and most commonly used components in modern radar and satellite communications systems [1-5]. While all-electronic methods are the traditional approaches used, they are subject to limitations in bandwidth. Photonic techniques are promising solutions for high performance RF filters and signal processors, offering wide bandwidths, high performance and versatility, and strong immunity to electromagnetic interference.

Hilbert transformers and temporal differentiators are both fundamental processing functions that are integral to many RF and microwave systems. They are quadrature filters that feature wideband 90° phase shifts, and are important techniques in the theory and practice of radar mapping, imaging edge detection as well as the realization of advanced modulation formats for digital communications. To further enhance the performance of these advanced signal process functions, a key approach is to realise fractional orders of these response functions to enable fractional Hilbert transformers, fractional differentiators, and potentially solving differential equations.

Optical fractional-order signal processors have been implemented based on fibre Bragg gratings [6-10], Mach-Zehnder interferometers [11], and ring-resonators [12]. Although these approaches offer many advantages they tend to suffer from limited bandwidth when processing signals.

Another approach is to use individual laser sources to supply the individual tap wavelengths. Better still, approaches that supply all of the tap wavelengths from a single source to generate high quality, and multiple wavelengths offer still further advantages. This can be achieved via mode-locked lasers [23], electro-optical modulation [24], and other approaches, and this provides solutions that have achieved attractive performance.

Optical frequency comb sources based on integrated nonlinear micro-resonators – Kerr microcombs – have recently been used for microwave signal processing [13-22]. They offer distinct advantages such as greatly reduced footprint and complexity and can provide a much higher number of wavelengths, thus offering fundamental advantages for the performance of photonic RF filters [25-32].

Here, we report a reconfigurable fractional microwave signal processor based on integrated microcombs. We report two kinds of signal processing functions, including fractional Hilbert transforms (FHT) and fractional differentiation (FD), achieved by programming and shaping the power of individual comb lines according to the required tap weights. System demonstrations of the FHT and FD are performed, including measurements of the RF amplitude and phase response, as well as the real-time impulse response for Gaussian input pulses.

## II. Theory

The FHT can be defined as the transfer function of a filter as follows:

$$H_P(\omega) = \begin{cases} e^{-j\varphi}, & if\ 0 \leq \omega < \pi \\ e^{j\varphi}, & if\ -\pi \leq \omega < 0 \end{cases} \quad (2.1)$$

where $\varphi = P \times \pi/2$ denotes the phase and P is the fractional order. As can be seen from (2.1), an FHT is a $\varphi$ phase shifter and the FHT becomes a classical HT when P = 1.

The impulse response of a fractional Hilbert transformer is a continuous hyperbolic function

$$h_p[t] = \begin{cases} \frac{1}{\pi \cdot t}, & t \neq 0 \\ \cot(\varphi), & t = 0 \end{cases} \quad (2.2)$$

which is truncated and sampled in time by discrete taps. The null frequency $f_c$ is determined by the sample spacing $\Delta t$

$$f_c = 1/\Delta t \quad (2.3)$$

In addition, the order of the FHT is continuously tunable by only adjusting the coefficient of the $9_{th}$ ($t$=0) tap while keeping the coefficients of other taps unchanged.

A temporal differentiator has a spectral transfer function that can be expressed as

$$H(\omega) = j(\omega - \omega_0)^N \quad (2.4)$$

where $\omega$ and $\omega_0$ are the angular frequencies of the input signal and the carrier, respectively, and N is an integer or fraction that reflects the order of the differentiator. For square root differentiation, $N = 1/2$ and a $\pi/4$ phase jump at $\omega_0$ is produced.

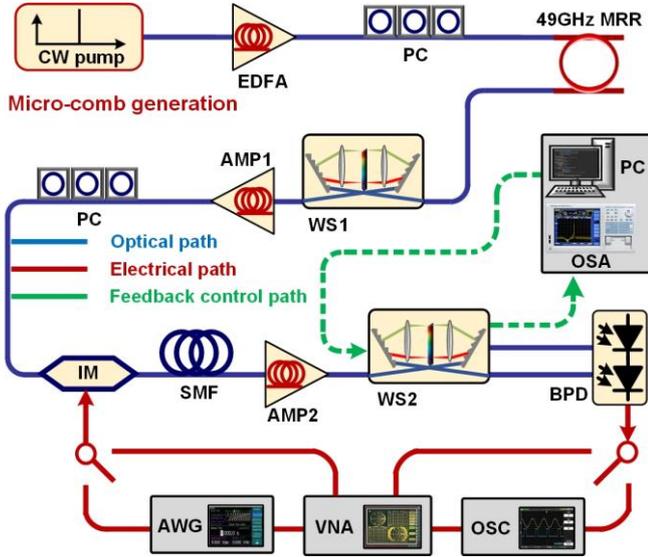

Fig. 1. Schematic diagram of the fractional Hilbert transformer based on an integrated 49GHz-spacing micro-comb source. EDFA: erbium-doped fiber amplifier. PC: polarization controller. MRR: micro-ring resonator. WSS: Wave shaper. IM: Intensity modulator. SMF: single mode fiber. OSA: optical spectrum analyzer. PC: computer. BPD: Balanced photodetector. VNA: vector network analyzer. AWG: arbitrary waveform generator. OSC: oscilloscope.

In order to implement these response functions, we use a powerful approach based on transversal filtering that has been extensively used to achieve a wide range of signal processing functions [29,30]. The transfer function of a photonic transversal filter is

$$H(\omega) = \sum_{n=0}^{N-1} h(n)e^{-j\omega nT} \quad (2.5)$$

where $\omega$ is the angular frequency of the input RF signal, $N$ is the number of taps, $h(n)$ is the discrete impulse response representing the tap coefficient of the nth taps, and $T$ is the time delay between adjacent taps. The free spectral range of the transversal filter $FSR_{RF}$ is given by $1/T$. By properly setting the tap coefficients ($h(n)$, $n = 0, 1, …, N$-1), virtually any response function or filter can be realized. Here, we demonstrate its capability for realizing both fractional Hilbert transformers and fractional order differentiators.

Figure 1 shows a schematic diagram of the fractional signal processor based on an integrated Kerr micro-comb source. It consists of two main blocks: one is a Kerr comb generation module based on a nonlinear MRR and another is a transversal filter module for reconfigurable fractional Hilbert transformer. Kerr optical frequency combs were generated in an integrated microring resonator (MRR), which was pumped by a continuous-wave (CW) laser. The CW pump was amplified by an erbium-doped fibre amplifier and the polarization adjusted via a polarization controller to optimize the power coupled to the MRR. When the pump wavelength was swept across one of the MRR's resonances and the pump power was high enough to provide sufficient parametric gain, optical parametric oscillation occurred, ultimately generating Kerr optical combs with a spacing equal to the free spectral range of the MRR. The generated Kerr micro-comb served as a multi-wavelength source where the power of each comb line was manipulated by Waveshapers to achieve the designed tap weights.

## III. EXPERIMENTAL RESULTS OF FHT

The integrated MRRs were fabricated on a high-index doped silica glass (n = ~1.7 at 1550 nm) platform using CMOS-compatible fabrication processes. The radii of the MRRs were both designed to be ~ 592 μm, corresponding to FSRs of ~0.4 nm (~49 GHz), which enabled a large number of comb wavelengths up to 80 in the C band and over 160 in the C+L band.

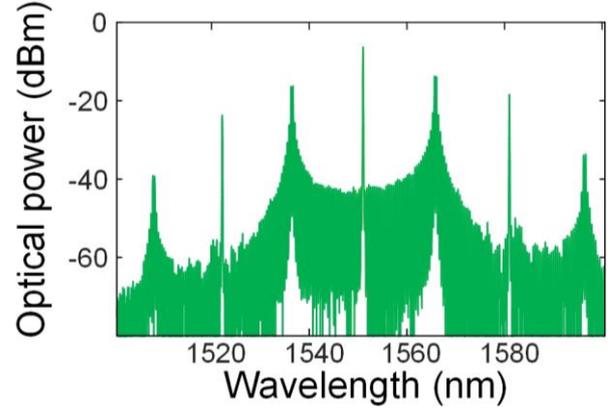

Fig. 2. Optical spectrum of the generated soliton crystal micro-comb.

To generate Kerr micro-combs, the pump power was set at ~30.5 dBm and the wavelength swept from blue to red. When the detuning between the pump wavelength and the MRR's cold resonance became small enough, such that the intra-cavity power reached a threshold value, modulation instability driven oscillation was initiated. As the detuning was changed further, distinctive 'fingerprint' optical spectra of soliton crystals were observed (Fig. 2) that are indicative of soliton crystals [19,29,30]. The comb was then spectrally shaped by two stages of Waveshapers (Finisar, 4000S) in order to enable a larger dynamic range of loss control and higher shaping accuracy than a single stage.

The micro-comb was first pre-shaped to reduce the power difference between the comb lines to under 5 dB. Next, the shaped comb lines were fed into an EO intensity modulator (IM) and then through ~2.1-km of standard single mode fibre (SMF) to provide a wavelength dependent delay. The dispersion of the SMF was 17.4 ps / (nm · km), corresponding to a time delay T of ~30 ps between adjacent taps, yielding an $FSR_{RF}$ of ~17 GHz for the fractional signal processor. The shaped comb lines were then amplified and accurately shaped by a Waveshaper according to the designed tap weights. For the second Waveshaper, a feedback control path was adopted to increase the accuracy of the comb shaping, where the power in the comb lines from one of the port of waveshaper were detected by an optical spectrum analyzer and then compared with the ideal tap weights in order to generate error signals for calibration. Finally, the weighted positive and negative taps out of the other two ports of the waveshaper were fed into both branches of the balanced photodetector. Since the Nyquist frequency was 24.5 GHz - in our case, half of the 49 GHz comb spacing - the operational bandwidth of the transversal filter was determined mainly by the $FSR_{RF}$ of 17 GHz.

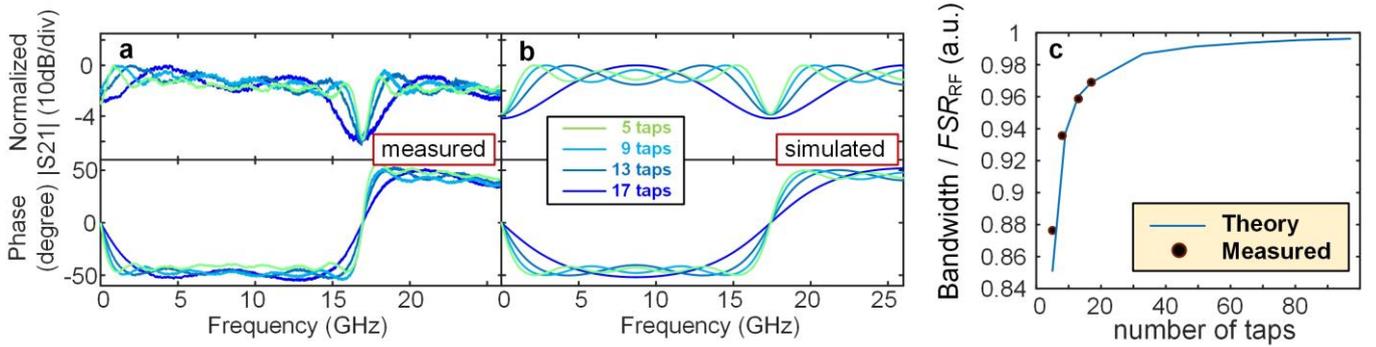

Fig. 3. Measured system RF frequency response for different number of filter taps based on a phase shift of 45⁰. (a) measured amplitude and phase response; (b) simulated amplitude and phase response; and (c) simulated and experimental result of 3 dB bandwidth with different taps.

The system RF frequency response was then characterized by using a vector network analyzer (VNA, Agilent MS4644B) to measure the system RF amplitude and phase frequency response. In Fig.3 (a), we measured the RF amplitude and phase frequency response of the fractional Hilbert transform filters with a 45 degree phase shift for 5, 9, 13 and 17 taps, respectively. Fig.3 (b) shows the theoretical results which all exhibit the expected behaviour. The theoretical normalized bandwidth / $FSR_{RF}$ as a function of the tap number is shown in Fig.3 (c). With a 17 tap filter, the FHT with a 45 degree phase shift exhibited a 3 dB bandwidth from 0.034 GHz to 16.45 GHz, corresponding to more than 8 octaves. Also, as can be seen, the theoretical BW increases with tap number, yielding a significant improvement in frequency selectivity.

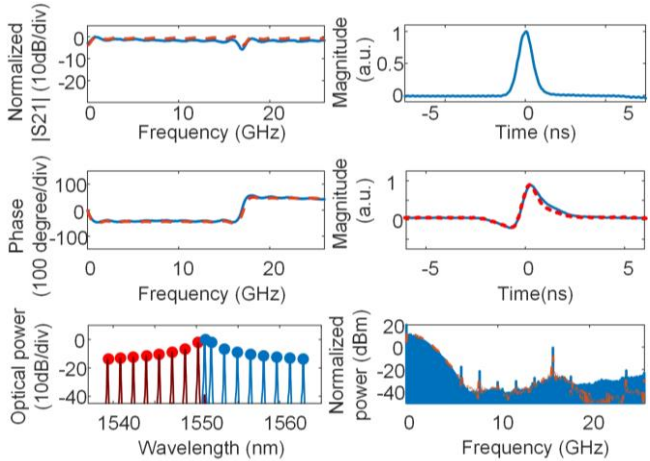

Fig. 4. Experimental results of 45⁰ phase shift fractional Hilbert transformer.

The input Gaussian pulses were generated from an arbitrary waveform generator (AWG, KEYSIGHT M9505A), shown in Figs. 4 and 5. As can be seen, the full-width at half-maximum is 50 ps with a bandwidth of 5 GHz. A good match between the power of the measured comb lines (red and blue solid lines) and the calculated ideal tap weights (red and blue dots) was obtained, indicating that the comb lines were successfully shaped.

The normalized frequency response and phase response of the FD with phase shift of 45 degree, corresponding to a tunable order of 0.5 is shown in fig. 5. We also performed systems demonstrations of real-time square rood differentiation for the Gaussian input pulses shown in Fig. 5. The measured output waveform agrees well with theory, further confirming the feasibility of our approach.

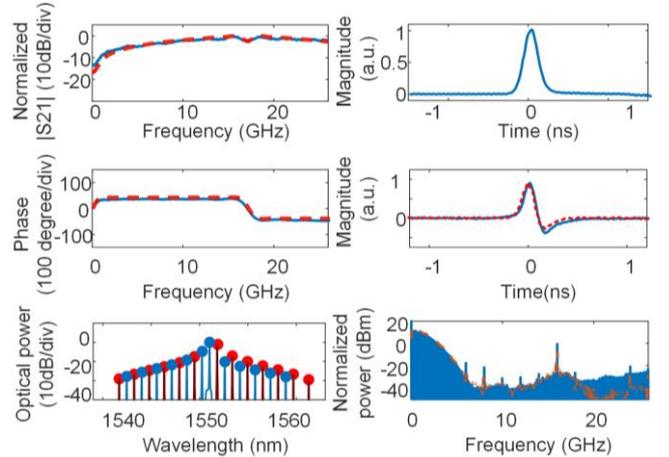

Fig. 5. Experimental results of optical intensity square root differentiator.

## IV. Conclusions

We demonstrate a reconfigurable microcomb-based fractional signal processor. The Kerr optical comb is produced via a CMOS-compatible nonlinear MRR, which greatly increases the processing bandwidth. By programming and shaping the individual comb lines' power according to the calculated taps weights, we successfully demonstrate a fractional Hilbert transformer and differentiator with an order of 0.5. The RF amplitude and phase response of the fractional signal processor are characterized, and system demonstrations of the real-time fractional signal processor are performed for Gaussian input pulses. The experimental results agree well with theory, verifying a promising new way to implement microwave photonic fractional signal processors featuring compact device footprint, high processing bandwidths and reconfigurability, for future ultra-high-speed microwave and computing and information processing systems.

ACKNOWLEDGMENT

This work was supported by the Australian Research Council Discovery Projects Program (No. DP150104327). RM acknowledges support by the Natural Sciences and Engineering Research Council of Canada (NSERC) through the Strategic, Discovery and Acceleration Grants Schemes, by the MESI PSR-SIIRI Initiative in Quebec, and by the Canada Research Chair Program. He also acknowledges additional support by the Government of the Russian Federation through the ITMO Fellowship and Professorship Program (grant 074-U 01) and by the 1000 Talents Sichuan Program in China. Brent E. Little was supported by the Strategic Priority

Research Program of the Chinese Academy of Sciences, Grant No. XDB24030000.

## REFERENCES

[1] J. Capmany, B. Ortega, and D. Pastor, "A tutorial on microwave photonic filters," *J. Lightwave Technol.,* vol. 24, no. 1, pp. 201-229, Jan. 2006.

[2] J. Capmany, and D. Novak, "Microwave photonics combines two worlds," *Nat. Photonics,* vol. 1, no. 6, pp. 319-330, Jun. 2007.

[3] J. P. Yao, "Microwave Photonics," *J. Lightwave Technol.,* vol. 27, no. 1-4, pp. 314-335, Jan-Feb. 2009.

[4] V. R. Supradeepa, C. M. Long, R. Wu, F. Ferdous, E. Hamidi, D. E. Leaird, and A. M. Weiner, "Comb-based radiofrequency photonic filters with rapid tunability and high selectivity," *Nat. Photonics,* vol. 6, no. 3, pp. 186-194, Mar. 2012.

[5] J. Wu, X. Xu, T. G. Nguyen, S. T. Chu, B. E. Little, R. Morandotti, A. Mitchell, and D. J. Moss, "RF Photonics: An Optical Microcombs' Perspective," *IEEE J. Sel. Top. Quantum Electron.,* vol. 24, no. 4, Jul-Aug. 2018.

[6] Li, M. and J. Yao, "Experimental Demonstration of a Wideband Photonic Temporal Hilbert Transformer Based on a Single Fiber Bragg Grating." *IEEE Photonics Technology Letters* vol. 22, no. 21, pp. 1559-1561, Nov. 2010.

[7] Asghari, M. H. and J. Azana, "All-optical Hilbert transformer based on a single phase-shifted fiber Bragg grating: design and analysis." *Opt Lett* vol. 34, no.3, pp. 334-336, Feb. 2009.

[8] Li, M. and J. Yao, "All-fiber temporal photonic fractional Hilbert transformer based on a directly designed fiber Bragg grating." *Opt Lett* vol. 35, no. 2, pp. 223-225, Jan. 2010.

[9] M. Delgado-Pinar, J. Mora, A. Diez, M. V. Andres, B. Ortega, and J. Capmany, "Tunable and reconfigurable microwave filter by use of a Bragg-grating-based acousto-optic superlattice modulator," *Opt. Lett.,* vol. 30, no. 1, pp. 8-10, Jan 1. 2005.

[10] G. Yu, W. Zhang, and J. A. R. Williams, "High-performance microwave transversal filter using fiber Bragg grating arrays," *IEEE Photonic Tech L,* vol. 12, no. 9, pp. 1183-1185, Sep. 2000.

[11] E. Hamidi, D. E. Leaird, and A. M. Weiner, "Tunable Programmable Microwave Photonic Filters Based on an Optical Frequency Comb," *IEEE Journal of Microwave Theory,* vol. 58, no. 11, pp. 3269-3278, Nov. 2010.

[12] Zhuang, L. M., et al, "Novel microwave photonic fractional Hilbert transformer using a ring resonator-based optical all-pass filter." *Optics Express* vol. 20, no. 24, pp. 26499-26510, Nov. 2012.

[13] P. Del'Haye, A. Schliesser, O. Arcizet, T. Wilken, R. Holzwarth, and T. J. Kippenberg, "Optical frequency comb generation from a monolithic microresonator," *Nature,* vol. 450, no. 7173, pp. 1214-1217, Dec 20. 2007.

[14] T. J. Kippenberg, R. Holzwarth, and S. A. Diddams, "Microresonator-Based Optical Frequency Combs," *Science,* vol. 332, no. 6029, pp. 555-559, Apr 29. 2011.

[15] L. Razzari, D. Duchesne, M. Ferrera, R. Morandotti, S. Chu, B. E. Little, and D. J. Moss, "CMOS-compatible integrated optical hyper-parametric oscillator," *Nat. Photonics,* vol. 4, no. 1, pp. 41-45, Jan 10. 2010.

[16] D. J. Moss, R. Morandotti, A. L. Gaeta, and M. Lipson, "New CMOS-compatible platforms based on silicon nitride and Hydex for nonlinear optics," *Nat. Photonics,* vol. 7, no. 8, pp. 597-607, Aug. 2013.

[17] J. S. Levy, A. Gondarenko, M. A. Foster, A. C. Turner-Foster, A. L. Gaeta, and M. Lipson, "CMOS-compatible multiple-wavelength oscillator for on-chip optical interconnects," *Nat. Photonics,* vol. 4, no. 1, pp. 37-40, Jan 10. 2010.

[18] A. Pasquazi, M. Peccianti, L. Razzari, D. J. Moss, S. Coen, M. Erkintalo, Y. K. Chembo, T. Hansson, S. Wabnitz, P. Del'Haye, X. X. Xue, A. M. Weiner, and R. Morandotti, "Micro-combs: A novel generation of optical sources," *Physics Reports,* vol. 729, pp. 1-81, Jan 27. 2018.

[19] D. C. Cole, E. S. Lamb, P. Del'Haye, S. A. Diddams, and S. B. Papp, "Soliton crystals in Kerr resonators," *Nat. Photonics,* vol. 11, no. 10, pp. 671-676, Oct. 2017.

[20] A. Pasquazi, L. Caspani, M. Peccianti, M. Clerici, M. Ferrera, L. Razzari, D. Duchesne, B. E. Little, S. T. Chu, D. J. Moss, and R. Morandotti, "Self-locked optical parametric oscillation in a CMOS compatible microring resonator: a route to robust optical frequency comb generation on a chip," *Opt. Express,* vol. 21, no. 11, pp. 13333-13341, Jun 3. 2013.

[21] A. Pasquazi, R. Ahmad, M. Rochette, M. Lamont, B. E. Little, S. T. Chu, R. Morandotti, and D. J. Moss, "All-optical wavelength conversion in an integrated ring resonator," *Opt. Express,* vol. 18, no. 4, pp. 3858-3863, Feb 15. 2010.

[22] H. Bao, A. Cooper, M. Rowley, L. Di Lauro, J. Sebastian T. Gongora, S. T. Chu, B. E. Little, G.L. Oppo, R. Morandotti, D. J. Moss, B. Wetzel, M. Peccianti and A. Pasquazi, "Laser Cavity-Soliton Micro-Combs", *Nature Photonics*, vol. 13, no. 6, 384–389, 2019.

[23] Ortigosa-Blanch, A., et al, "Tunable radio-frequency photonic filter based on an actively mode-locked fiber laser." *Opt Lett* vol. 35, no. 6, pp. 709-711, Mar. 2006.

[24] Supradeepa, V. R., et al. "Comb-based radiofrequency photonic filters with rapid tunability and high selectivity." *Nature Photonics* vol. 6, no. 3, pp. 186-194, Mar. 2012.

[25] T. G. Nguyen, M. Shoeiby, S. T. Chu, B. E. Little, R. Morandotti, A. Mitchell, and D. J. Moss, "Integrated frequency comb source-based Hilbert transformer for wideband microwave photonic phase analysis," *Opt. Express,* vol. 23, no. 17, pp. 22087-22097, Aug 24. 2015.

[26] X. Xu, M. Tan, J. Wu, T. G. Nguyen, S. T. Chu, B. E. Little, R. Morandotti, A. Mitchell, and D. J. Moss, "Advanced Adaptive Photonic RF Filters with 80 Taps Based on an Integrated Optical Micro-Comb Source," *Journal of Lightwave Technology,* vol. 37, no. 4, pp. 1288-1295, Feb. 2019.

[27] X. X. Xue, Y. Xuan, H. J. Kim, J. Wang, D. E. Leaird, M. H. Qi, and A. M. Weiner, "Programmable Single-Bandpass Photonic RF Filter Based on Kerr Comb from a Microring," *J. Lightwave Technol.,* vol. 32, no. 20, pp. 3557-3565, Oct 15. 2014.

[28] X. Xu, J. Wu, M. Shoeiby, T. G. Nguyen, S. T. Chu, B. E. Little, R. Morandotti, A. Mitchell, and D. J. Moss, "Reconfigurable broadband microwave photonic intensity differentiator based on an integrated optical frequency comb source," *APL Photonics,* vol. 2, no. 9, 096104, Sep. 2017. doi: 10.1063/1.4989871.

[29] X. Xu, J. Wu, T. G. Nguyen, T. Moein, S. T. Chu, B. E. Little, R. Morandotti, A. Mitchell, and D. J. Moss, "Photonic microwave true time delays for phased array antennas using a 49 GHz FSR integrated optical micro-comb source [Invited]," *Photonics Res,* vol. 6, no. 5, pp. B30-B36, May 1. 2018.

[30] X. Xu, M. Tan, J. Wu, T. G. Nguyen, S. T. Chu, B. E. Little, R. Morandotti, A. Mitchell, and D. J. Moss, "High performance RF filters via bandwidth scaling with Kerr micro-combs," *APL Photonics,* vol. 4, no. 2, pp. 026102. 2019.

[31] X. Xu, J. Wu, T. G. Nguyen, S. Chu, B. Little, A. Mitchell, R. Morandotti, and D. J. Moss, "Broadband RF Channelizer based on an Integrated Optical Frequency Kerr Comb Source," *Journal of Lightwave Technology,* vol. 36, no. 19, 4519-4526 (2018).
DOI: 10.1109/JLT.2018.2819172. 2018.

[32] X. Xu, J. Wu, T. G. Nguyen, M. Shoeiby, S. T. Chu, B. E. Little, R. Morandotti, A. Mitchell, and D. J. Moss, "Advanced RF and microwave functions based on an integrated optical frequency comb source," *Opt. Express,* vol. 26, no. 3, pp. 2569-2583, Feb 5. 2018.